%% file: main.tex
\documentclass[sigconf]{acmart}

\AtBeginDocument{%
  \providecommand\BibTeX{{%
    \normalfont B\kern-0.5em{\scshape i\kern-0.25em b}\kern-0.8em\TeX}}}

\setcopyright{acmcopyright}
\copyrightyear{2022}
\acmYear{2022}
\acmDOI{XXXXXXX.XXXXXXX}

\acmConference[CHI '23]{CHI 2022}{April 23--28,
  2023}{Hamburg, Germany}
\acmPrice{15.00}
\acmISBN{978-1-4503-XXXX-X/18/06}

\usepackage{pdfpages}

\begin{document}

\title{Using dynamic circles and squares to visualize spatio-temporal variation}


\author{Harsh Patel}
\authornote{Both authors contributed equally to this research.}
\email{hpatel01@terpmail.umd.edu}
\author{Nicole Schneider}
\authornotemark[1]
\email{nsch@umd.edu}
\author{Hanan Samet}
\email{hjs@cs.umd.edu}
\affiliation{%
  \institution{University of Maryland}
  \city{College Park}
  \state{Maryland}
  \country{USA}
  \postcode{20742}
}

\renewcommand{\shortauthors}{Patel and Schneider et al.}

\begin{abstract}
  Visualizations such as bar charts, scatter plots, and objects on geographical maps often convey critical information, including exact and relative numeric values, using shapes. The choice of shape and method of encoding information is often arbitrarily, or based on convention. However, past studies have shown that the human eye can be fooled by visual representations. The Ebbinghaus illusion demonstrates that the perceived relative sizes of shapes depends on their configuration, which in turn can affect judgements, especially in visualizations like proportional symbol maps. In this study we evaluate the effects of varying the type of shapes and metrics for encoding data in visual representations on a spatio-temporal map interface. We find that some combinations of shape and metric are more conducive to accurate human judgements than others, and provide recommendations for applying these findings in future visualization designs. 
\end{abstract}




\maketitle

\section{Introduction} \label{intro}
\input{intro}

\section{Related work} \label{related}
\input{related}

\section{CoronaViz System} \label{coronaviz}
\input{coronaviz}

\section{Methods} \label{methods}
\input{methods}

\section{Results} \label{results}
\input{results}

\section{Discussion} \label{discussion}
\input{discussion}

\section{Future Work} \label{future}
\input{future}

\section{Conclusion} \label{conclusion}
\input{conclusion}


\begin{acks}
Special thanks to Terry Slocum and Brian Ondov for their input.
\end{acks}

\bibliographystyle{ACM-Reference-Format}
\bibliography{main}


\appendix
\section{Survey} \label{survey_appendix}
The entire survey is available here: \href{https://forms.gle/RaL3HgH6MxoqQx7d6}{Form} 

\end{document}

%% file: intro.tex
Visualizations such as bar charts, scatter plots, and objects on geographical maps often convey critical information, including exact and relative numeric values, using shapes. Given a dataset, there are many valid choices that can be made about how to present the information, but there are relatively few definitive conclusions about what shapes and metrics should be used to promote accurate perception of the data shown in the visualization. In fact, the choice of shape and method of encoding information is often chosen arbitrarily, or based on convention. However, past studies have shown that the human eye can be fooled by visual representations. The Ebbinghaus illusion demonstrates that the perceived relative sizes of shapes depends on their configuration, which in turn can affect judgements, especially in visualizations like proportional symbol maps. 

We present a study that evaluates the effects of varying the type of shape and metric for encoding data in visual representations on a spatio-temporal map interface. We use real data encoded in various ways on a production data visualization and exploration system for tracking COVID-19 related statistics through space and time. We perform a user study to determine which combinations of symbol shape, encoding metric, and type of variation targeted (spatial or temporal) promote the most accurate perceptions in the context of this system. We measure participant responses to a survey of multiple choice questions and analyze the results to draw conclusions about the conditions under which people make more or less accurate judgements of the relative sizes of shapes on a map visualization.

With respect to encoding metrics, we hypothesize that metrics requiring little mental manipulation, such as diameter of a circle, will yield better participant performance than metrics requiring substantial mental manipulation, such as circumference of a circle. Intuitively, we are positing that because diameter simply consists of estimating the distance between two points on the circle, participants will be able to make this estimation more accurately than they can do for more complex metrics like perimeter or area. Regarding spatio-temporal variation, we hypothesize that questions depicting spatial variation on a single map will yield more accurate estimates of relative size than questions depicting temporal variation using side by side map snapshots, which require participants to look back and forth between two separate maps with identical background configurations.

Our main contributions are the findings we show that indicate some combinations of shape and metric are more conducive to accurate human judgements than others. We also detail our methodology which can easily be extended to test different shapes and metrics, or new attributes altogether, depending on the context under study. We also provide recommendations for applying our findings in future map visualization design.

The rest of the paper is organized as follows. In Section \ref{related} we present a review of previous work. We then describe the system used to contextualize our research questions in Section \ref{coronaviz}, and our methodology in Section \ref{methods}. Finally, we present results (Section \ref{results}), a discussion of the findings (Section \ref{discussion}), avenues for future work (Section \ref{future}), and conclusions (Section \ref{conclusion}).

%% file: related.tex
\subsection{Visualization and Visual Perception}
Visualizations are a common method for representing data in an easily digestible manner. Data can be encoded in any number of ways, which varies depending on the type of visualization and the type of data. However, all visualizations rely on visual perception as a key underlying principle.

\paragraph{Visual Perception}
There is a sizeable body of work dealing with visual perception and visualizations, including perceptual correspondence between data and its visualization \cite{Dast02}. Healy et al.\ \cite{Heal12} gives a survey of visual attention and memory, explaining principles related to the visual system, and what it sees and misses in different scenarios. Heer et al.\ \cite{Heer10} shows that Amazon Mechanical Turk (MTurk), the same crowdsourcing platform we use in this work, is a viable way to conduct many visualization perception studies.

\paragraph{Illusions in Visualizations}
When presenting data in visual form, one must be cognizant of the phenomenon of visual illusions. There are many such illusions, including the widely studied Ebbinghaus Illusion, where the perceived size of a circle can be influenced by the surrounding circles, as well as a number of other factors \cite{Mass71}. Visual illusions have been studied under a variety of conditions in the psychology literature \cite{Dewi15}, as well as in the data visualization field \cite{Hong22, Mitt14} and more recently in the context of virtual reality (VR) \cite{Horo03, Egeb21}. In the latter, characteristics of data are represented visually using shape, surface properties, and motion through VR. This can lead to illusions in how geometric structures are perceived due to their properties in the VR world \cite{Horo03}. From the data visualization perspective, there is recent work \cite{Hong22} detailing the systematic bias in tri-variate scatter plots, when encoding a third dimension of information in size or color. Visual perception is shown to be sensitive to choice in size or color range, which leads to misjudgements. All of this work serves as motivation for studying which kinds of objects and object properties lead to the most accurate visual perceptions in data visualization.

\subsection{Spatial, Temporal, and Cartographic Visualizations}
Many works have focused on developing or modifying spatial visualizations in novel ways, to allow for easier human perception \cite{Beck95, Droc11}. In particular, Drocourt et al.\ \cite{Droc11} develop an algorithm for visualizing the advancement/retreat of glaciers in Greenland using radial lines and nested rings. They use a nonlinear mapping to generate angular coordinates from Cartesian coordinates, which allows for consistent spatial perception. This work represents a unique use of circles in spatial visualization, where arc length (a segment of the circumference of a circle) conveys vital information. In our work, we find that the circumference of a circle is one of the most visually challenging metrics for participants to estimate accurately on maps (out of the 3 metrics and 2 shapes we tested across both spatial and temporal questions).

Temporal visualizations can be constructed in a number of different ways \cite{Daas05}, but often treat time as an additional axis, or include animation or interactivity to convey changes over time. Examples include \cite{Hao04}, which develops an interactive display of large molecule datasets in biology. In our study, the system we use to generate the images we present to participants is spatio-temporal, meaning it incorporates aspects of both spatial and temporal visualization. We further describe the context in which we compare spacial and temporal visualization queries in Section \ref{map_questions}.

Cartographic visualization has evolved substantially with the rise of modern post-computing mapping \cite{MacE13}. In particular, the rapid development of mapping applications spawned the study of cartographic interaction, which is the subject of \cite{Roth11}. The link between scientific visualization and cartographic visualization is thoroughly discussed in \cite{Mace97, Fair01}.

\subsection{Role of Shapes and Encoding Metrics in Map Visualization}
One of the key aspects of cartographic visualization that we test in our study is choice of shape. We know that the observable size of a circle can be influenced by factors such as the size difference between a target circle and adjacent circles in a close proximity \cite{Gilm81}. However, many visualizations use circles as the primary shape to represent data, especially in a geographical context \cite{Meih73, Clev82}. There are several works that study the perception of shapes in a geographical context \cite{Stac18, Sloc09, Craw73}. Many of these works indicate that choice of shape is important, in addition to other factors like background and dimensionality of encoding metric which may also influence perception. Stachon et al.\ \cite{Stac18}, studies the effect of shape (circle vs. triangle) on the speed of processing when a map background is present and not present. Another study \cite{Craw73} finds that graduated squares built on the basis of area rather than a linear dimension were estimated accurately. Groop et al.\ \cite{Groo78} finds that overlap of circles also affects the perception of relative sizes and proposes transparent overlap rather than partial occlusion to help combat these effects.
Legend values have also been explored as a way to improve the effectiveness of a geographic display that uses circles to represent data \cite{Dobs74}. However, if done incorrectly, the legend values can inhibit the conversion of information. Cox et al.\ assesses the efficacy of value scaling against the use of numerous legend symbols for both circles and squares \cite{Cox76}. Results showed that the use of various legend symbols on a map yielded more correct shape estimations compared to the apparent value scaling. 

\subsection{Map Animation}
Although map animation is outside the scope of our study, it is a common way to convey temporal (and non-temporal) geographic data. Early work on map animation to represent geographic-scale change was done by \cite{Harr02}. See also \cite{Sloc22} for an in-depth analysis of geovisualization and the evolution of map animation. Studies have also devised methods to improve comprehension of multivariate geographic data \cite{Turk14, Dorl92} and make animated maps easier to comprehend by reducing cognitive overload \cite{Lucj16}.


%% file: coronaviz.tex
CoronaViz \footnote{\url{https://coronaviz.umiacs.io/}} \footnote{\url{https://coronaviz.umiacs.io/squares/}} \cite{Same20} is a dynamic COVID-19 disease visualization system that was created in light of the coronavirus pandemic as a new way to track and visualize pandemic-related data over time. The system displays various data including confirmed cases, deaths, recoveries, hospitalizations, positivity rate, etc.\ on a single interactive and multi-layer geographical display. The data displayed corresponds to specific locations on a map that allows the user to select, hover, zoom in or out, and pan. This system builds on many of the key principles of interactive map interfaces described in \cite{Teit08, Sank09}, and differs from many of the COVID-19 reporting visualizations \cite{toolJHU, toolNYT} in that it supports zooming which increases the resolution of data presented, as additional smaller units become visible. 

The graphical interface for CoronaViz represents data using hollow circles which we call `geocircles' whose radii are determined by the values of the variables they represent. Animation control buttons allow users to search through time manually or view the data unfolding in accelerated time, giving a summary of the temporal changes in the data. However, in order for users to glean an accurate picture of pandemic status and progression through time, users must be able to accurately estimate the relative size of a geocircle, given the other geocircles visible on the map, as well as the size of the same geocircle at a different (no longer visible) point in time. This raises the main question of our study- what shape and encoding metric should be used to allow for the most accurate perception of numerical values or relative numerical values on a dynamic map interface like CoronaViz?

%% file: methods.tex
\subsection{Description of the Survey}
The survey is made up of 9 parts. Part 1 and 2 consist of the consent form and a question asking the user to provide their MTurk Worker ID. Parts 3, 4, and 5 consist of the 24 main survey questions, which are multiple choice style. These questions present the participant with one or two map visualizations and ask them to estimate, by eye, the relative sizes of shapes on the map(s). These questions are described in full detail in Section \ref{map_questions}. There is also an attention check question mixed in with the aforementioned 24 questions. The attention check is used to decide which responses are good faith attempts, and which are the result of random guessing (we discard these responses). More details about the content of this question are given in Section \ref{attention_question}. Part 6 contains 2 questions which inquire participants' opinions about which shape types and metrics they found easiest to estimate. Part 7 contains 2 Ebbinghaus Illusion questions, and Parts 8 and 9 record demographic info and provide a survey completion code. A link to the entire survey is given in Appendix \ref{survey_appendix}.

\subsubsection{Map Query Questions} \label{map_questions}
The map query questions comprise the majority of the survey. Each of these questions provide the participant with one or two images of the CoronaViz map interface, which includes several shapes representing COVID-19 statistics by location. However, the underlying statistics that drive the sizes of the shapes are hidden from the participants. Instead, the only numerical value visible on the map is a label for the \emph{reference shape}. The participants are also told which metric was used to encode this value (area, circumference/perimeter, or diameter/side length). Using those two pieces of information, the participants are asked to visually estimate the relative size of a second shape on the map, termed the \emph{query shape}, which is labeled with a question mark '?'.
Each map query question is designed to evaluate the participant's ability to estimate the relative size of shapes in situations that vary across several attributes of interest, which are described in further detail below.

\begin{figure}[ht]
  \centering
  \includegraphics[width=1.0\linewidth]{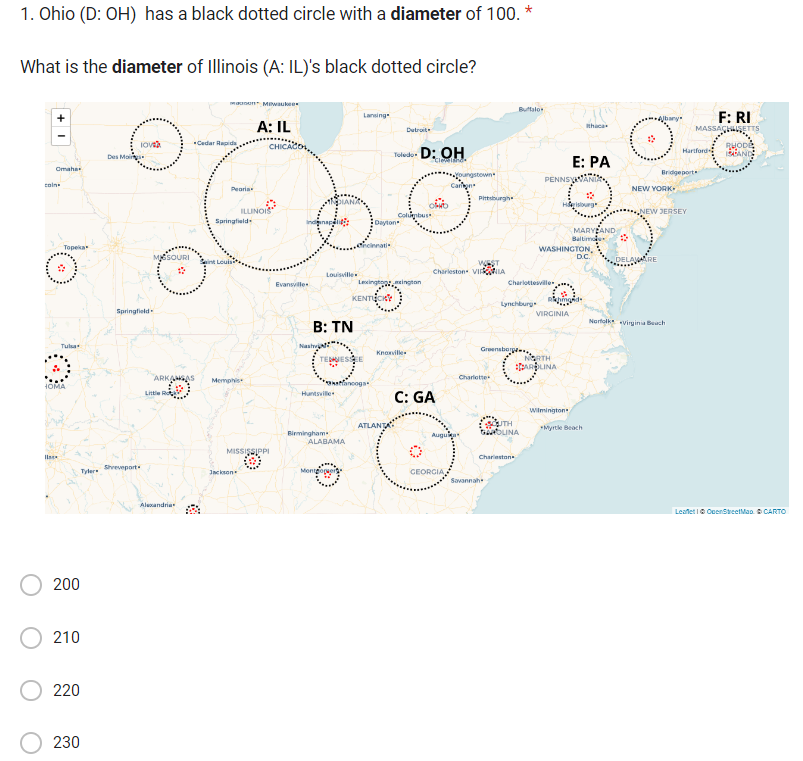}
  \caption{Map query question using a circle to encode diameter across spatial variation.}
   \label{fig:circle_diameter_spatial}
\end{figure}

\paragraph{Spatial Questions}
All of the map query questions that we presented to the participants fell into one of two categories, spatially-focused questions and temporally-focused questions. The first type (termed spatial questions), present the participant with scenarios where they must estimate the size of a shape in one location, given a reference shape in a \emph{different location} on the same map. An example of one such question is given in Figure \ref{fig:circle_diameter_spatial}. These questions measure the participant's ability to visually estimate variation in shape size across space.

\begin{figure}[ht]
  \centering
  \includegraphics[width=1.0\linewidth]{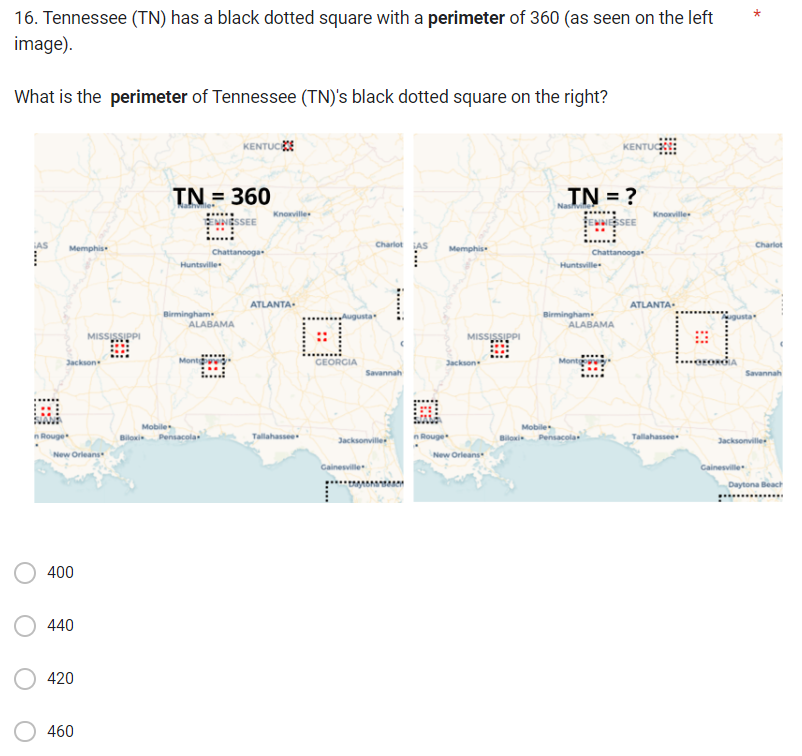}
  \caption{Map query question using a square to encode perimeter across temporal variation.}
   \label{fig:square_perim_temporal}
\end{figure}

\paragraph{Temporal Questions}
The other type of map query questions, temporal questions, present the participant with two distinct maps that have identical background perspectives (Figure \ref{fig:square_perim_temporal}). These two maps represent two snapshots of one location undergoing temporal animation. In other words, we use static side-by-side images that, when considered in tandem, convey a temporal variation in shape size. This design choice allows us to isolate an important difference between spatial and temporal variation: the fact that temporal variation consists of a single center point, around which a shape is changing size, whereas spatial variation consists of both changing shape size, and changing center location.

\begin{figure}[ht]
  \centering
  \includegraphics[width=1.0\linewidth]{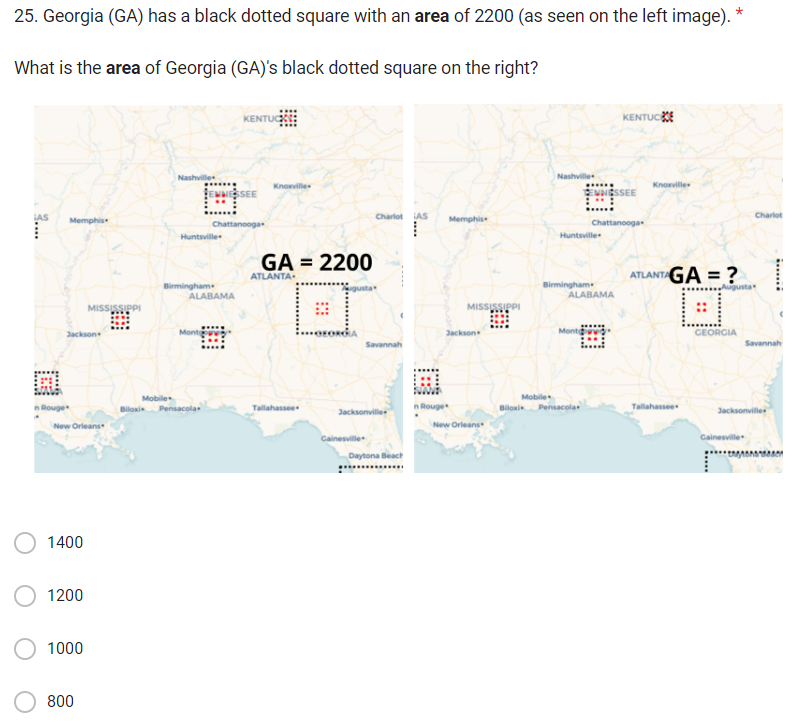}
  \caption{Map query question of using a square to encode area across temporal variation.}
   \label{fig:square_area_temporal}
\end{figure}

\paragraph{Shapes}
To represent the numeric values associated with different locations on the map, each of the map query questions presents one of two types of shapes, whose sizes are scaled according to their encoding metric (see Section \ref{encoding}). These values are either presented using \emph{circles} (ex. Figure \ref{fig:circle_diameter_spatial}) or \emph{squares} (ex. Figure \ref{fig:square_area_temporal}). Circles are the canonical choice for map symbols \cite{Meih73, Clev82}, but squares have been shown to allow for good visual estimation under certain conditions \cite{Craw73}.  

\paragraph{Encoding Metric} \label{encoding}
We also vary the metric used to encode the numerical values for the shapes presented in our survey. For each shape type, we encode numbers using diameter/side length (ex. Figure \ref{fig:circle_diameter_spatial}), circumference/perimeter (ex. Figure \ref{fig:square_perim_temporal}), and area (ex. Figure) \ref{fig:square_area_temporal}. 

\subsubsection{Attention Check} \label{attention_question}
\begin{figure}[ht]
  \centering
  \includegraphics[width=1.0\linewidth]{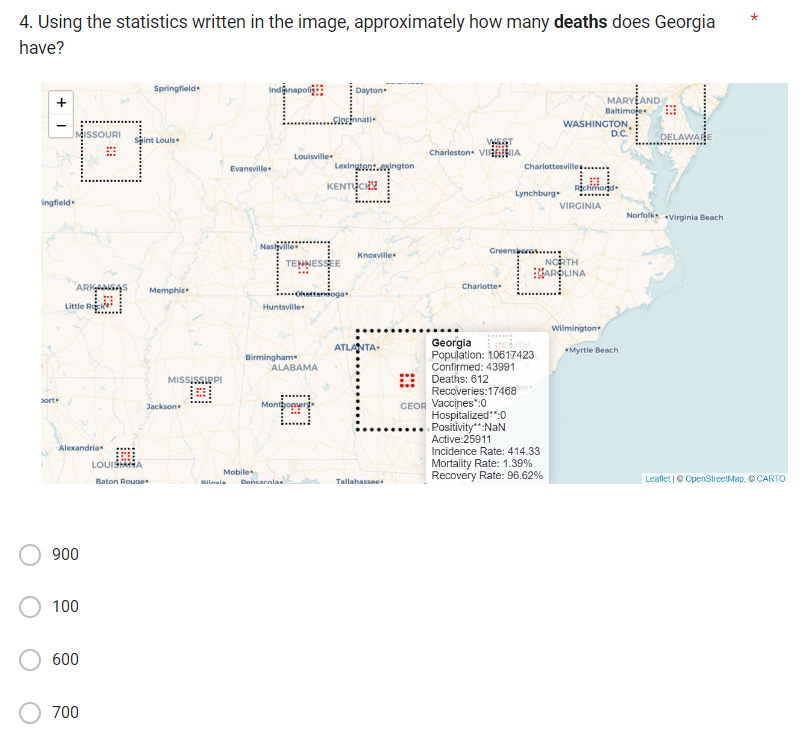}
  \caption{Attention check question used to discard survey responses made via random guessing. The question has an obvious correct answer of 600.}
   \label{fig:attention_check}
\end{figure}

Disguised within the map query questions is one attention check question (Figure \ref{fig:attention_check}). This question is designed so that the correct answer is easily derivable from the information presented in the summary box overlaid on the map image. We include this question to separate out responses in which the participant randomly guessed from those where the participant made a good faith attempt to read and answer each question. We eliminate from consideration the entire survey response for any participant who did not answer the attention check question correctly.

\subsubsection{Other Questions}

\paragraph{Opinion Questions}
After completing the map query questions we ask participants to provide their opinions about which shape and encoding metric they found easiest to estimate. These questions are designed to gather participant feedback that can be directly compared to their actual success rates for the map query questions, to determine how well the attitudes about shapes and encoding metrics align with the actual performance across these attributes.

\begin{figure}[ht]
  \centering
  \includegraphics[width=1.0\linewidth]{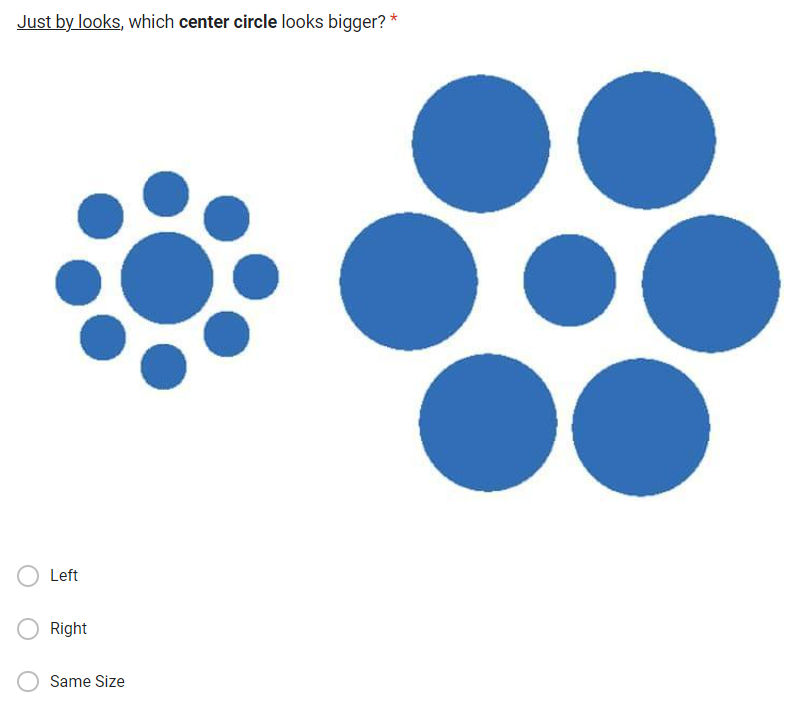}
  \caption{Question demonstrating the typical Ebbinghaus illusion with circles as the objects.}
   \label{fig:circle_illusion}
\end{figure}

\begin{figure}[ht]
  \centering
  \includegraphics[width=1.0\linewidth]{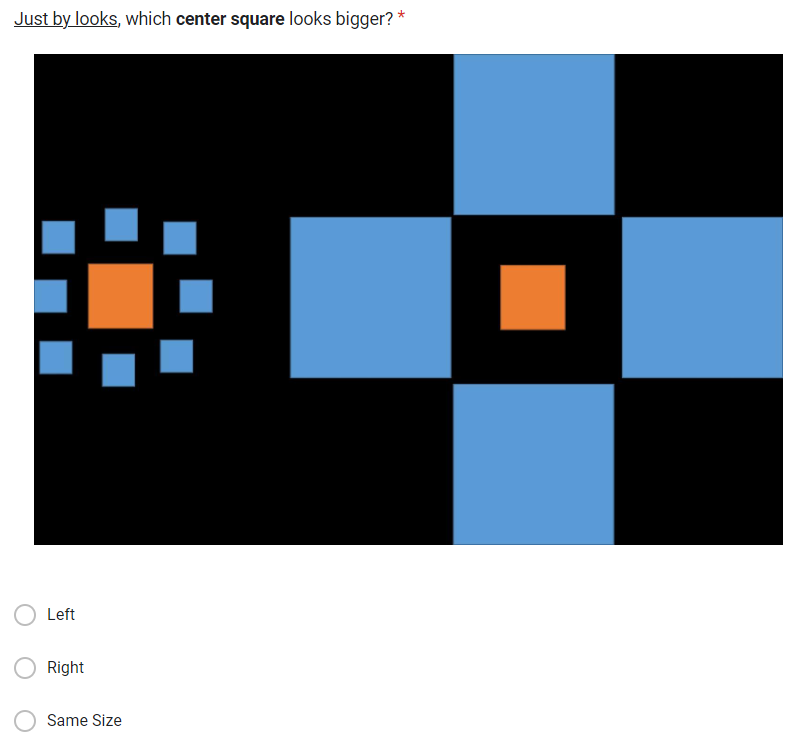}
  \caption{Question designed to demonstrate the Ebbinghaus illusion using squares as the objects instead of circles.}
   \label{fig:square_illusion}
\end{figure}

\paragraph{Illusion Questions}
Next we show two questions (Figures \ref{fig:circle_illusion} and \ref{fig:square_illusion}) containing images demonstrating the Ebbinghaus Illusion for circles \footnote{\url{https://www.theguardian.com/science/head-quarters/2016/aug/22/the-ebbinghaus-illusion-small-far-away-circles-father-ted}} and squares respectively, and ask participants to estimate which center shape is larger in each case.

\paragraph{Demographic Questions}
Finally, the participants are asked to provide optional demographic information. This includes a question for gender, highest degree completed or in progress, and age range.

\subsection{Recruiting Participants}
We recruited 24 participants for the study using the Amazon Mechanical Turk (MTurk) crowdsource platform. We limited the task to only allow crowdworkers located in the United States. When workers accepted our task, they used the survey link provided in the task to access our survey (see Appendix \ref{survey_appendix}). After answering all of the questions in the survey, workers received a completion code that they then provided via the MTurk interface to complete the job. We paid workers \$2.50 for completing the task, which is a target of about \$8-10 per hour based on the number of questions in the survey and our estimates for the time it would take to complete the survey.

\subsection{Selecting Reference and Query Values}

We chose the reference and query values for the questions so that the two questions per condition (i.e. the two questions for Circle, Area, Spatial) cover both the case where the reference value is larger than the query value and also the case where the reference value is smaller than the query value. We did so to combat any effects due changes in difficulty of estimating a larger value given a smaller one, versus estimating a smaller value given a larger one. 

We also chose values appropriately sized to the task. For instance, values encoded with the Area metric were larger across the board, so that the overall sizes of the shapes stayed relatively similar across all questions. While we allowed some variation in shape size which is natural within the CoronaViz platform, we ensured that no reference or query shape consisted of more than approximately one quarter of the map background, to help combat any effects that may arise out of difficulty estimating very large shapes on the map.

\subsection{Selecting Reference and Query Positions}
We recognize that distance between reference and query shapes may impact the difficulty of estimating their relative sizes. For temporal questions, the reference and query shapes were presented in two identical side by side map backgrounds. This means the distance between the centers of these two shapes is constant across all temporal questions, since the map sizes are held constant from question to question. In our survey this distance was approximately 1040 pixels. Of note, the nature of the two maps side by side requires a small visual break between the maps, which in turn means that the distances for temporal questions were higher than for spatial questions. We discuss the implications of this in Section \ref{discussion}.

For spatial questions, it is more difficult to keep a consistent distance between reference and query shapes, while also maintaining a variety of locations to combat learning effects from question to question. We settle on a middle ground by varying the locations on the map, but ensuring that all distances from reference to query shape (measured center to center) are between 200 and 900 pixels. Further, for any one condition tested, there are always two questions for which the scores are averaged. We ensure that the average distance for between reference and query for any condition is between 450 and 600 pixels.

\subsection{Setting Multiple Choice Options}
For simplicity we rounded all correct answers to the nearest multiple of 10, and chose incorrect (distractors) that were also multiples of 10. We used a pixel ruler \footnote{\url{https://www.rapidtables.com/web/tools/pixel-ruler.html}} to measure the actual diameter or side length of each reference and query shape, and used that to mathematically calculate circumference/perimeter or area if applicable for the question. To account for human error in measuring the values, we used the pixel ruler five times for each question computed the average before rounding to the nearest 10. We also scaled the answers and distractors down to a range we thought participants could reasonably be asked to estimate: at or below 4000 for area, 2000 for perimeter/circumference, and 600 for diameter/side length. 

%% file: results.tex
We collected responses from twenty-four participants, of which we retained and report results on the sixteen responses that passed the attention check question. We summarize the overall results from parts 3, 4, and 5 of the survey in Tables \ref{table:overall_summary} and \ref{table:spatiotemporal_summary}.

\subsection{Opinion Questions} \label{opinions}
Out of the sixteen participants, seven said circles were the easiest to estimate, five said squares were easiest, and the remaining four said they were of equal difficulty. For the encoding metrics, six participants found circumference/perimeter to be the easiest, five said diameter/side length was easiest, one thought area was easiest, and four found them to be of the same difficulty.

\subsection{Analysis of Map Query Questions}
A summary of the overall performance is presented in Table \ref{table:overall_summary} and a summary of the performance for each disjoint combination is presented in Table \ref{table:spatiotemporal_summary}. 

Each participant answered two questions for each combination of shape, metric, and spatial/temporal question type. This means that for each scenario (such as Circle-Area-Spatial or Circle-Diameter-Temporal) a participant could have answered 0, 1, or 2 of the questions correctly. We aggregate these to determine the number of correct responses per participant for each of the tests we perform (Circle vs. Square, Spatial vs. Temporal, etc.). 

For all statistical tests we apply the Bonferroni correction \cite{Hoch90} to adjust the significance level required to reject the null hypothesis, since we are performing multiple hypothesis tests on the same dataset. Rather than testing all possible combinations of shape, metric, and question type, we select a few based on the RQs outlined in Section \ref{intro} and the participant feedback discussed in Section \ref{opinions}.

We first perform the Shapiro-Wilk test for normality for each pair of scenarios we test, which showed in each case that the data was not normally distributed (p < 0.05). As a result, we used the non-parametric test Wilcoxen Signed Rank test to test each of the following hypothesis.

\paragraph{Shape}

In our first test we aimed to find out if squares are easier to estimate than circles. For each participant we count the number of correctly answered questions which used the circle shape, and then number of correctly answered questions which used the square shape. We then use the Wilcoxen Signed Rank test to determine if the median difference is zero (null hypothesis) or if it is not zero (alternative hypothesis). We discard the ties and find p > $\alpha$, indicating no significant difference in median between the two groups.

\paragraph{Spatial/Temporal}
In the next test we aimed to find out if spatial type questions are easier to estimate than temporal ones. For each participant we count the number of correctly answered spatial and temporal questions and use the one tailed Wilcoxen Signed Rank test to determine if the median difference between scores for spatial and temporal questions is zero or greater than zero. We discard the ties, and find that $p < \alpha$, meaning that spatial type questions are significantly easier than temporal type questions.

\paragraph{Encoding Metric}
In the opinion questions we found that participants reported having the easiest time estimating circumference/perimeter and diameter/side length for metrics and circle for shape. As such, we test to see if one of these two metrics is easier to estimate accurately for circles. We again use the Wilcoxen Signed Rank test to determine if the median difference between scores for Circle-Diameter and Circle-Perimeter questions is zero or greater than zero. We discard the ties and find that $p < \alpha$, meaning that the diameter of a circle is significantly easier to estimate than perimeter of a circle.

\subsection{Demographics}
The majority of the participants, eleven, where between the ages of 21-30. Three were between the ages of 31-40 and two were between 41-50. Half of the participants were male and the other half are female. We had one individual with a high school degree or equivalent, three who had an associates degree, seven with a bachelors degree, and five who hold a masters or professional degree. 

\subsection{Illusion Questions}   
The circle illusion question (Figure \ref{fig:circle_illusion}) garnered a correct response (circles are the same size) from one participant. Of the incorrect responses, twelve participants said the left middle circle was larger and three indicated the right one was larger. For the square illusion question (Figure \ref{fig:square_illusion}), four participants correctly said that the center squares were the same size, nine thought the left center square was larger and three thought the right was larger. 

\subsection{Summary}
Based on the results of our three tests and the opinions of the participants, it is clear that no particular shape or metric is easier to estimate across the board. We found that squares were slightly (but not statistically significantly) easier to estimate than circles. On the other hand, the opinion questions indicated that more participants thought circles were easier to estimate than squares. This potential disconnect between what participants think is easier to estimate and what they are better at estimating in practice is an interesting avenue of future study. For spatial and temporal questions, we found that the spatial questions were significantly easier to estimate than temporal questions, supporting our hypothesis that the visual separation and extra distance between the snapshots provided in the temporal questions made them more difficult than the spatial questions, which presented only a single map to look at. For circle questions in particular, we observed that participants estimated diameter significantly more accurately than they estimated perimeter. This is also in line with our hypothesis that metrics requiring less mental manipulation, like diameter, would be easier to estimate than metrics requiring more complex manipulations, like unfolding the circumference of the circle and estimating its length.

\begin{table}
\begin{tabular}{ |p{2.5cm}|p{3cm}|p{2.3cm}|  }
 \hline
 \multicolumn{3}{|c|}{Overall Summary Statistics} \\
 \hline
  & & Success Rate (\%)\\
 \hline
 Shape type & Circle & 25.5  \\
 & Square & \textbf{28.1} \\
 \hline
 Encoding metric & Diameter/Side length & \textbf{29.7} \\
 & Perimeter & 21.1 \\
 & Area & \textbf{29.7} \\
 \hline
 Variation & Spatial & \textbf{32.3} \\
 & Temporal & 21.4 \\
 \hline
\end{tabular}
\caption{Overall summary of participant performance across attributes of interest for the sixteen participants who passed the attention check question. Note these attributes are not disjoint.}
\label{table:overall_summary}
\end{table}

\begin{table}
\begin{tabular}{ |p{3cm}|p{1cm}|p{1cm}|p{1cm}|p{1cm}|  }
 \hline
 & \multicolumn{2}{|c|}{Spatial Results} & \multicolumn{2}{c|}{Temporal Results} \\
 \hline
  & Circle & Square & Circle & Square \\
 \hline
 Diameter/Side length & 40.6 & 37.5 & 25.0  & 15.6\\
 Perimeter & 15.6 & 43.8 & 12.5 &  12.5\\
 Area & 37.5 & 18.8 & 21.9 & 40.6\\
 \hline
\end{tabular}
\caption{Success rate (\%) of participants estimating shapes across 3 attributes of interest. Results are given as the average over 2 questions for each combination of attributes. There are 12 disjoint combinations tested.}
\label{table:spatiotemporal_summary}
\end{table}

%% file: discussion.tex
Looking at the overall results, squares yield slightly higher performance than circles, diameter/side length and area have better performance than perimeter, and spatial questions have better performance than temporal ones.
Looking at the disjoint combinations, we see that side length and perimeter perform similarly for squares in both spatial and temporal questions. This fits with the intuition that for squares, perimeter is simply a 4x multiple of side length, which should be just as easy (or difficult) to estimate. No such simple relationship exists for circles, which show more mixed results depending on the encoding metric and question type.

\subsection{Limitations}
For this study we surveyed 25 participants using MTurk. However, with 9 participants failing the attention check question, we only retained 16 responses to use in analysis. With this relatively small sample size, we were only able to find significant differences between a few combinations of shape, metric, and spatial/temporal question type.

By design, our survey is also limited in its ability to test temporal queries in particular. We chose to design the temporal question to test one aspect of temporal changes, the change in shape size while center point holds steady, and ignore other aspects that make temporal questions challenging. This includes the need to remember, rather than reference, the previous representation. In our study participants could look back and forth between one snapshot and another to estimate the difference. However, even with this advantage, we found that performance on spatial questions was significantly better than on the temporal snapshots. We attribute this to the added distance between the query and reference shape for the temporal questions, which came about as a result of presenting two maps side by side with a small visual break between them. A future study could be designed to incorporate animation, which better captures the complexities of temporal queries.

We also consider that by deploying the survey on a crowdsource platform like MTurk, we have no control over the resolution of the screens used while taking the survey. This is an inherent limitation to all visualization studies deployed in this manner, and is discussed extensively in \cite{Heer10}.

Finally, in designing the questions, we made trade-offs with respect to allowing or controlling variation of distance between the reference and query point. We decided to keep the distances within some reasonable bounds, rather than allowing complete variation. Ideally, the distances should be held constant from question to question to eliminate possible confounding effects, but this undermines the natural variability intrinsic to a real system like CoronaViz. Since we did notice that temporal questions led to significantly worse performance than spatial ones, we suggest as future work a study that explicitly measures the effects of reference-query distance in a map setting like ours.

%% file: future.tex
There are a few avenues of future work that we believe would enhance the results presented in this study. One aspect of the CoronaViz interface that we did not address directly here is the presentation of multiple metrics per location using concentric shapes of different colors. It would be interesting to study how well people are able to estimate the relative sizes of the outer and inner shape, to determine if this is indeed a useful way to convey multiple data values per location.

For the purposes of this study, we used static images taken from a graphical interface. This gave the user a visual to compare the encoded mark to another one at all times. To account for this, future studies can be conducted where a user is shown an animation or a GIF from the graphical interface and asked to determine how the mark changed over time instead of having the original reference object to look at indefinitely. This would account for the temporal queries. As for the spatial questions, a user might be asked to compare the growth or shrinkage of particular mark that is either spatially close or distant. Both would test how well a user can memorize the original size of the mark and use their memory as a reference instead of staring at it on a screen. Studies where data are encoded in other scale, such as log or linear, must also be considered and compared with one another as data values may exceed the human eyes limitation to preserve visualizations. In general, however, further research and extensive studies should be conducted to find what shapes and metrics allow for the most accurate visuals created from data. Data encoded as circles by convention may not be the best in all scenarios.

%% file: conclusion.tex
Previous work has shown that visual perception can be influenced by a number of factors, including the type of shape being viewed and the background it is viewed on. With this in mind, we studied how well people are able to visually estimate the relative sizes of different shapes in cartographic visualizations taken from a real system for visualizing COVID-19 data. We varied the choice of shape, metric by which numbers are encoded visually, and type of variation depicted across the reference and query shapes- either spatial variation or temporal variation (via side by side snapshots). We found that when using circles as the visualization shape, diameter was significantly easier to estimate than circumference. We also found that participants more accurately estimated relative sizes for spatial queries than for temporal ones, which we believe is attributable to the increased distance between reference and query object in the temporal question setup. Ultimately, we have shown that choice of shape and metric makes a real difference in how map visualizations are perceived by viewers. As a result, we hope that these findings spur further research along the lines we have suggested and encourage scientists as well as cartographers to consider carefully how they present numerical data in map visualizations moving forward.

%% file: main.bbl

\begin{thebibliography}{37}


\ifx \showCODEN    \undefined \def \showCODEN     #1{\unskip}     \fi
\ifx \showDOI      \undefined \def \showDOI       #1{#1}\fi
\ifx \showISBNx    \undefined \def \showISBNx     #1{\unskip}     \fi
\ifx \showISBNxiii \undefined \def \showISBNxiii  #1{\unskip}     \fi
\ifx \showISSN     \undefined \def \showISSN      #1{\unskip}     \fi
\ifx \showLCCN     \undefined \def \showLCCN      #1{\unskip}     \fi
\ifx \shownote     \undefined \def \shownote      #1{#1}          \fi
\ifx \showarticletitle \undefined \def \showarticletitle #1{#1}   \fi
\ifx \showURL      \undefined \def \showURL       {\relax}        \fi
\providecommand\bibfield[2]{#2}
\providecommand\bibinfo[2]{#2}
\providecommand\natexlab[1]{#1}
\providecommand\showeprint[2][]{arXiv:#2}

\bibitem[Hor(2003)]%
        {Horo03}
 \bibinfo{year}{2003}\natexlab{}.
\newblock \bibinfo{booktitle}{\emph{{Looking Just Like It Doesn’t: Perception
  and Illusion in Scientific Visualization}}}. \bibinfo{series}{Fluids
  Engineering Division Summer Meeting}, Vol.~\bibinfo{volume}{Volume 2:
  Symposia, Parts A, B, and C}.
\newblock
\urldef\tempurl%
\url{https://doi.org/10.1115/FEDSM2003-45198}
\showDOI{\tempurl}
\showeprint{https://asmedigitalcollection.asme.org/FEDSM/proceedings-pdf/FEDSM2003/36975/1629/2593161/1629\_1.pdf}


\bibitem[Becker et~al\mbox{.}(1995)]%
        {Beck95}
\bibfield{author}{\bibinfo{person}{Richard~A. Becker},
  \bibinfo{person}{Stephen~G. Eick}, {and} \bibinfo{person}{Allan~R. Wilks}.}
  \bibinfo{year}{1995}\natexlab{}.
\newblock \showarticletitle{Visualizing Network Data}.
\newblock \bibinfo{journal}{\emph{IEEE Trans. Vis. Comput. Graph.}}
  \bibinfo{volume}{1} (\bibinfo{year}{1995}), \bibinfo{pages}{16--28}.
\newblock


\bibitem[Cleveland et~al\mbox{.}(1982)]%
        {Clev82}
\bibfield{author}{\bibinfo{person}{William~S. Cleveland},
  \bibinfo{person}{Charles~S. Harris}, {and} \bibinfo{person}{Robert McGill}.}
  \bibinfo{year}{1982}\natexlab{}.
\newblock \showarticletitle{Judgments of circle sizes on statistical maps}.
\newblock \bibinfo{journal}{\emph{J. Amer. Statist. Assoc.}}
  \bibinfo{volume}{77}, \bibinfo{number}{379} (\bibinfo{year}{1982}),
  \bibinfo{pages}{541--547}.
\newblock
\urldef\tempurl%
\url{https://doi.org/10.1080/01621459.1982.10477844}
\showDOI{\tempurl}


\bibitem[Cox(1976)]%
        {Cox76}
\bibfield{author}{\bibinfo{person}{Carleton~W. Cox}.}
  \bibinfo{year}{1976}\natexlab{}.
\newblock \showarticletitle{Anchor effects and the estimation of graduated
  circles and squares}.
\newblock \bibinfo{journal}{\emph{The American Cartographer}}
  \bibinfo{volume}{3}, \bibinfo{number}{1} (\bibinfo{year}{1976}),
  \bibinfo{pages}{65--74}.
\newblock
\urldef\tempurl%
\url{https://doi.org/10.1559/152304076784080195}
\showDOI{\tempurl}


\bibitem[Crawford(1973)]%
        {Craw73}
\bibfield{author}{\bibinfo{person}{Paul~V. Crawford}.}
  \bibinfo{year}{1973}\natexlab{}.
\newblock \showarticletitle{The perception of graduated squares as cartographic
  symbols}.
\newblock \bibinfo{journal}{\emph{The Cartographic Journal}}
  \bibinfo{volume}{10}, \bibinfo{number}{2} (\bibinfo{year}{1973}),
  \bibinfo{pages}{85--88}.
\newblock
\urldef\tempurl%
\url{https://doi.org/10.1179/caj.1973.10.2.85}
\showDOI{\tempurl}


\bibitem[Daassi et~al\mbox{.}(2005)]%
        {Daas05}
\bibfield{author}{\bibinfo{person}{Chaouki Daassi}, \bibinfo{person}{Laurence
  Nigay}, {and} \bibinfo{person}{Marie-Christine Fauvet}.}
  \bibinfo{year}{2005}\natexlab{}.
\newblock \showarticletitle{A taxonomy of temporal data visualization
  techniques}.
\newblock \bibinfo{journal}{\emph{Information-Interaction-Intelligence}}
  \bibinfo{volume}{5}, \bibinfo{number}{2} (\bibinfo{year}{2005}),
  \bibinfo{pages}{41--63}.
\newblock


\bibitem[Dastani(2002)]%
        {Dast02}
\bibfield{author}{\bibinfo{person}{Mehdi Dastani}.}
  \bibinfo{year}{2002}\natexlab{}.
\newblock \showarticletitle{The Role of Visual Perception in Data
  Visualization}.
\newblock \bibinfo{journal}{\emph{Journal of Visual Languages \& Computing}}
  \bibinfo{volume}{13}, \bibinfo{number}{6} (\bibinfo{year}{2002}),
  \bibinfo{pages}{601--622}.
\newblock
\showISSN{1045-926X}
\urldef\tempurl%
\url{https://doi.org/10.1006/jvlc.2002.0235}
\showDOI{\tempurl}


\bibitem[{de Wit} et~al\mbox{.}(2015)]%
        {Dewi15}
\bibfield{author}{\bibinfo{person}{Matthieu~M. {de Wit}}, \bibinfo{person}{John
  {van der Kamp}}, {and} \bibinfo{person}{Rob Withagen}.}
  \bibinfo{year}{2015}\natexlab{}.
\newblock \showarticletitle{Visual illusions and direct perception: Elaborating
  on Gibson's insights}.
\newblock \bibinfo{journal}{\emph{New Ideas in Psychology}}
  \bibinfo{volume}{36} (\bibinfo{year}{2015}), \bibinfo{pages}{1--9}.
\newblock
\showISSN{0732-118X}
\urldef\tempurl%
\url{https://doi.org/10.1016/j.newideapsych.2014.07.001}
\showDOI{\tempurl}


\bibitem[Dobson(1974)]%
        {Dobs74}
\bibfield{author}{\bibinfo{person}{Micheal~W Dobson}.}
  \bibinfo{year}{1974}\natexlab{}.
\newblock \showarticletitle{Refining legend values for proportional circle
  maps}.
\newblock \bibinfo{journal}{\emph{Cartographica: The International Journal for
  Geographic Information and Geovisualization}} \bibinfo{volume}{11},
  \bibinfo{number}{1} (\bibinfo{year}{1974}), \bibinfo{pages}{45–53}.
\newblock
\urldef\tempurl%
\url{https://doi.org/10.3138/ktw4-7562-6811-8h21}
\showDOI{\tempurl}


\bibitem[Dorling(1992)]%
        {Dorl92}
\bibfield{author}{\bibinfo{person}{Daniel Dorling}.}
  \bibinfo{year}{1992}\natexlab{}.
\newblock \showarticletitle{Stretching Space and Splicing Time: From
  Cartographic Animation to Interactive Visualization}.
\newblock \bibinfo{journal}{\emph{Cartography and Geographic Information
  Systems}} \bibinfo{volume}{19}, \bibinfo{number}{4} (\bibinfo{year}{1992}),
  \bibinfo{pages}{215--227}.
\newblock
\urldef\tempurl%
\url{https://doi.org/10.1559/152304092783721259}
\showDOI{\tempurl}
\showeprint{https://doi.org/10.1559/152304092783721259}


\bibitem[Drocourt et~al\mbox{.}(2011)]%
        {Droc11}
\bibfield{author}{\bibinfo{person}{Yoann Drocourt}, \bibinfo{person}{Rita
  Borgo}, \bibinfo{person}{Kilian Scharrer}, \bibinfo{person}{Tavi Murray},
  \bibinfo{person}{Suzanne Bevan}, {and} \bibinfo{person}{Min Chen}.}
  \bibinfo{year}{2011}\natexlab{}.
\newblock \showarticletitle{Temporal Visualization of Boundary-based
  Geo-information Using Radial Projection}.
\newblock \bibinfo{journal}{\emph{Comput. Graph. Forum}}  \bibinfo{volume}{30}
  (\bibinfo{date}{06} \bibinfo{year}{2011}), \bibinfo{pages}{981--990}.
\newblock
\urldef\tempurl%
\url{https://doi.org/10.1111/j.1467-8659.2011.01947.x}
\showDOI{\tempurl}


\bibitem[Egeberg et~al\mbox{.}(2021)]%
        {Egeb21}
\bibfield{author}{\bibinfo{person}{Mie Egeberg}, \bibinfo{person}{Stine Lind},
  \bibinfo{person}{Niels~C. Nilsson}, {and} \bibinfo{person}{Stefania
  Serafin}.} \bibinfo{year}{2021}\natexlab{}.
\newblock \showarticletitle{Exploring the Effects of Actuator Configuration and
  Visual Stimuli on Cutaneous Rabbit Illusions in Virtual Reality}. In
  \bibinfo{booktitle}{\emph{ACM Symposium on Applied Perception 2021}} (Virtual
  Event, France) \emph{(\bibinfo{series}{SAP '21})}.
  \bibinfo{publisher}{Association for Computing Machinery},
  \bibinfo{address}{New York, NY, USA}, Article \bibinfo{articleno}{1},
  \bibinfo{numpages}{9}~pages.
\newblock
\showISBNx{9781450386630}
\urldef\tempurl%
\url{https://doi.org/10.1145/3474451.3476230}
\showDOI{\tempurl}


\bibitem[Fairbairn et~al\mbox{.}(2001)]%
        {Fair01}
\bibfield{author}{\bibinfo{person}{David Fairbairn}, \bibinfo{person}{Gennady
  Andrienko}, \bibinfo{person}{Natalia Andrienko}, \bibinfo{person}{Gerd
  Buziek}, {and} \bibinfo{person}{Jason Dykes}.}
  \bibinfo{year}{2001}\natexlab{}.
\newblock \showarticletitle{Representation and its Relationship with
  Cartographic Visualization}.
\newblock \bibinfo{journal}{\emph{Cartography and Geographic Information
  Science}} \bibinfo{volume}{28}, \bibinfo{number}{1} (\bibinfo{year}{2001}),
  \bibinfo{pages}{13--28}.
\newblock
\urldef\tempurl%
\url{https://doi.org/10.1559/152304001782174005}
\showDOI{\tempurl}


\bibitem[Gilmartin(1981)]%
        {Gilm81}
\bibfield{author}{\bibinfo{person}{Patricia~P. Gilmartin}.}
  \bibinfo{year}{1981}\natexlab{}.
\newblock \showarticletitle{Influences of map context on circle perception}.
\newblock \bibinfo{journal}{\emph{Annals of the Association of American
  Geographers}} \bibinfo{volume}{71}, \bibinfo{number}{2}
  (\bibinfo{year}{1981}), \bibinfo{pages}{253--258}.
\newblock
\urldef\tempurl%
\url{https://doi.org/10.1111/j.1467-8306.1981.tb01351.x}
\showDOI{\tempurl}


\bibitem[Groop and Cole(1978)]%
        {Groo78}
\bibfield{author}{\bibinfo{person}{Richard~E Groop} {and}
  \bibinfo{person}{Daniel Cole}.} \bibinfo{year}{1978}\natexlab{}.
\newblock \showarticletitle{Overlapping graduated circles / magnitude
  estimation and method of portrayal}.
\newblock \bibinfo{journal}{\emph{Cartographica: The International Journal for
  Geographic Information and Geovisualization}} \bibinfo{volume}{15},
  \bibinfo{number}{2} (\bibinfo{year}{1978}), \bibinfo{pages}{114–122}.
\newblock
\urldef\tempurl%
\url{https://doi.org/10.3138/q5q5-n244-8462-ng25}
\showDOI{\tempurl}


\bibitem[Hao et~al\mbox{.}(2004)]%
        {Hao04}
\bibfield{author}{\bibinfo{person}{Xuejun Hao}, \bibinfo{person}{Amitabh
  Varshney}, {and} \bibinfo{person}{Sergei Sukharev}.}
  \bibinfo{year}{2004}\natexlab{}.
\newblock \showarticletitle{Real-Time Visualization of Large Time-Varying
  Molecules}.
\newblock  (\bibinfo{date}{08} \bibinfo{year}{2004}).
\newblock


\bibitem[Harrower(2002)]%
        {Harr02}
\bibfield{author}{\bibinfo{person}{Mark~A. Harrower}.}
  \bibinfo{year}{2002}\natexlab{}.
\newblock \emph{\bibinfo{title}{Visual benchmarks: Representing geographic
  change with map animation}}.
\newblock \bibinfo{thesistype}{Ph.\,D. Dissertation}.
\newblock
\showISBNx{978-0-493-84599-9}
\urldef\tempurl%
\url{https://www.proquest.com/dissertations-theses/visual-benchmarks-representing-geographic-change/docview/275797248/se-2}
\showURL{%
\tempurl}
\newblock
\shownote{Copyright - Database copyright ProQuest LLC; ProQuest does not claim
  copyright in the individual underlying works; Last updated - 2022-01-18}.


\bibitem[Healey and Enns(2012)]%
        {Heal12}
\bibfield{author}{\bibinfo{person}{Christopher Healey} {and}
  \bibinfo{person}{James Enns}.} \bibinfo{year}{2012}\natexlab{}.
\newblock \showarticletitle{Attention and Visual Memory in Visualization and
  Computer Graphics}.
\newblock \bibinfo{journal}{\emph{IEEE Transactions on Visualization and
  Computer Graphics}} \bibinfo{volume}{18}, \bibinfo{number}{7}
  (\bibinfo{year}{2012}), \bibinfo{pages}{1170--1188}.
\newblock
\urldef\tempurl%
\url{https://doi.org/10.1109/TVCG.2011.127}
\showDOI{\tempurl}


\bibitem[Heer and Bostock(2010)]%
        {Heer10}
\bibfield{author}{\bibinfo{person}{Jeffrey Heer} {and} \bibinfo{person}{Michael
  Bostock}.} \bibinfo{year}{2010}\natexlab{}.
\newblock \showarticletitle{Crowdsourcing graphical perception: using
  mechanical turk to assess visualization design} \emph{(\bibinfo{series}{CHI
  '10})}. \bibinfo{publisher}{Association for Computing Machinery},
  \bibinfo{address}{New York, NY, USA}, \bibinfo{pages}{203–212}.
\newblock
\showISBNx{9781605589299}
\urldef\tempurl%
\url{https://doi.org/10.1145/1753326.1753357}
\showDOI{\tempurl}


\bibitem[Hochberg and Benjamini(1990)]%
        {Hoch90}
\bibfield{author}{\bibinfo{person}{Yosef Hochberg} {and} \bibinfo{person}{Yoav
  Benjamini}.} \bibinfo{year}{1990}\natexlab{}.
\newblock \showarticletitle{More powerful procedures for multiple significance
  testing}.
\newblock \bibinfo{journal}{\emph{Statistics in medicine}} \bibinfo{volume}{9},
  \bibinfo{number}{7} (\bibinfo{year}{1990}), \bibinfo{pages}{811--818}.
\newblock


\bibitem[Hong et~al\mbox{.}(2022)]%
        {Hong22}
\bibfield{author}{\bibinfo{person}{Matt-Heun Hong}, \bibinfo{person}{Jessica~K.
  Witt}, {and} \bibinfo{person}{Danielle~Albers Szafir}.}
  \bibinfo{year}{2022}\natexlab{}.
\newblock \showarticletitle{The Weighted Average Illusion: Biases in Perceived
  Mean Position in Scatterplots}.
\newblock \bibinfo{journal}{\emph{IEEE Transactions on Visualization and
  Computer Graphics}} \bibinfo{volume}{28}, \bibinfo{number}{1}
  (\bibinfo{year}{2022}), \bibinfo{pages}{987--997}.
\newblock
\urldef\tempurl%
\url{https://doi.org/10.1109/TVCG.2021.3114783}
\showDOI{\tempurl}


\bibitem[Howard et~al\mbox{.}(2008)]%
        {Sloc09}
\bibfield{author}{\bibinfo{person}{H. Howard}, \bibinfo{person}{R. McMaster},
  \bibinfo{person}{T. Slocum}, {and} \bibinfo{person}{F. Kessler}.}
  \bibinfo{year}{2008}\natexlab{}.
\newblock \showarticletitle{Thematic cartography and geovisualization}.
\newblock  (\bibinfo{year}{2008}).
\newblock


\bibitem[MacEachren and Taylor(2013)]%
        {MacE13}
\bibfield{author}{\bibinfo{person}{A.M. MacEachren} {and}
  \bibinfo{person}{D.R.F. Taylor}.} \bibinfo{year}{2013}\natexlab{}.
\newblock \bibinfo{booktitle}{\emph{Visualization in Modern Cartography}}.
\newblock \bibinfo{publisher}{Elsevier Science}.
\newblock
\showISBNx{9781483287928}
\urldef\tempurl%
\url{https://books.google.com/books?id=3cP-BAAAQBAJ}
\showURL{%
\tempurl}


\bibitem[Maceachren and Kraak(1997)]%
        {Mace97}
\bibfield{author}{\bibinfo{person}{Alan~M. Maceachren} {and}
  \bibinfo{person}{Menno-Jan Kraak}.} \bibinfo{year}{1997}\natexlab{}.
\newblock \showarticletitle{Exploratory cartographic visualization: Advancing
  the agenda}.
\newblock \bibinfo{journal}{\emph{Computers \& Geosciences}}
  \bibinfo{volume}{23}, \bibinfo{number}{4} (\bibinfo{year}{1997}),
  \bibinfo{pages}{335--343}.
\newblock
\showISSN{0098-3004}
\urldef\tempurl%
\url{https://doi.org/10.1016/S0098-3004(97)00018-6}
\showDOI{\tempurl}
\newblock
\shownote{Exploratory Cartograpic Visualisation}.


\bibitem[Massaro and Anderson(1971)]%
        {Mass71}
\bibfield{author}{\bibinfo{person}{Dom Massaro} {and} \bibinfo{person}{Norman
  Anderson}.} \bibinfo{year}{1971}\natexlab{}.
\newblock \showarticletitle{Judgmental model of the Ebbinghaus illusion}.
\newblock \bibinfo{journal}{\emph{Journal of experimental psychology}}
  \bibinfo{volume}{89} (\bibinfo{date}{08} \bibinfo{year}{1971}),
  \bibinfo{pages}{147--51}.
\newblock
\urldef\tempurl%
\url{https://doi.org/10.1037/h0031158}
\showDOI{\tempurl}


\bibitem[Meihoefer(1973)]%
        {Meih73}
\bibfield{author}{\bibinfo{person}{Hans-Joachim Meihoefer}.}
  \bibinfo{year}{1973}\natexlab{}.
\newblock \showarticletitle{The visual perception of the circle in thematic
  maps/experimental results}.
\newblock \bibinfo{journal}{\emph{Cartographica: The International Journal for
  Geographic Information and Geovisualization}} \bibinfo{volume}{10},
  \bibinfo{number}{1} (\bibinfo{year}{1973}), \bibinfo{pages}{63–84}.
\newblock
\urldef\tempurl%
\url{https://doi.org/10.3138/2771-5577-5417-369t}
\showDOI{\tempurl}


\bibitem[Mittelstädt et~al\mbox{.}(2014)]%
        {Mitt14}
\bibfield{author}{\bibinfo{person}{Sebastian Mittelstädt},
  \bibinfo{person}{Andreas Stoffel}, {and} \bibinfo{person}{Daniel Keim}.}
  \bibinfo{year}{2014}\natexlab{}.
\newblock \showarticletitle{Methods for Compensating Contrast Effects in
  Information Visualization}.
\newblock \bibinfo{journal}{\emph{Computer Graphics Forum}}
  \bibinfo{volume}{33}.
\newblock
\urldef\tempurl%
\url{https://doi.org/10.1111/cgf.12379}
\showDOI{\tempurl}


\bibitem[Roth(2011)]%
        {Roth11}
\bibfield{author}{\bibinfo{person}{Robert Roth}.}
  \bibinfo{year}{2011}\natexlab{}.
\newblock \emph{\bibinfo{title}{Interacting With Maps: The Science And Practice
  Of Cartographic Interaction}}.
\newblock \bibinfo{thesistype}{Ph.\,D. Dissertation}.
\newblock


\bibitem[Samet et~al\mbox{.}(2020)]%
        {Same20}
\bibfield{author}{\bibinfo{person}{Hanan Samet}, \bibinfo{person}{Yunheng Han},
  \bibinfo{person}{John Kastner}, {and} \bibinfo{person}{Hong Wei}.}
  \bibinfo{year}{2020}\natexlab{}.
\newblock \showarticletitle{Using animation to visualize spatio-temporal
  varying {COVID-19} data}. In \bibinfo{booktitle}{\emph{Proceedings of the ACM
  SIGSPATIAL Workshop on Modeling and Understanding the Spread of COVID-19}}
  \emph{(\bibinfo{series}{COVID-19})}. \bibinfo{publisher}{Association for
  Computing Machinery}, \bibinfo{address}{New York, NY, USA},
  \bibinfo{pages}{53--62}.
\newblock
\urldef\tempurl%
\url{https://doi.org/10.1145/3423459.3430761}
\showDOI{\tempurl}


\bibitem[Sankaranarayanan et~al\mbox{.}(2009)]%
        {Sank09}
\bibfield{author}{\bibinfo{person}{Jagan Sankaranarayanan},
  \bibinfo{person}{Hanan Samet}, \bibinfo{person}{Benjamin~E. Teitler},
  \bibinfo{person}{Michael~D. Lieberman}, {and} \bibinfo{person}{Jon
  Sperling}.} \bibinfo{year}{2009}\natexlab{}.
\newblock \showarticletitle{TwitterStand: News in Tweets}
  \emph{(\bibinfo{series}{GIS '09})}. \bibinfo{publisher}{Association for
  Computing Machinery}, \bibinfo{address}{New York, NY, USA},
  \bibinfo{pages}{42–51}.
\newblock
\showISBNx{9781605586496}
\urldef\tempurl%
\url{https://doi.org/10.1145/1653771.1653781}
\showDOI{\tempurl}


\bibitem[Slocum et~al\mbox{.}(2022)]%
        {Sloc22}
\bibfield{author}{\bibinfo{person}{Terry~A Slocum}, \bibinfo{person}{Robert~B
  McMaster}, \bibinfo{person}{Fritz~C Kessler}, {and} \bibinfo{person}{Hugh~H
  Howard}.} \bibinfo{year}{2022}\natexlab{}.
\newblock \bibinfo{booktitle}{\emph{Thematic cartography and
  geovisualization}}.
\newblock \bibinfo{publisher}{CRC Press}.
\newblock


\bibitem[Stachoň et~al\mbox{.}(2018)]%
        {Stac18}
\bibfield{author}{\bibinfo{person}{Zdeněk Stachoň}, \bibinfo{person}{Čeněk
  Šašinka}, \bibinfo{person}{Jiří Čeněk}, \bibinfo{person}{Stephan
  Angsüsser}, \bibinfo{person}{Petr Kubíček}, \bibinfo{person}{Zbyněk
  Štěrba}, {and} \bibinfo{person}{Martina Bilíková}.}
  \bibinfo{year}{2018}\natexlab{}.
\newblock \showarticletitle{Effect of Size, Shape and Map Background in
  Cartographic Visualization: Experimental Study on Czech and Chinese
  Populations}.
\newblock \bibinfo{journal}{\emph{ISPRS International Journal of
  Geo-Information}} \bibinfo{volume}{7}, \bibinfo{number}{11}
  (\bibinfo{year}{2018}).
\newblock
\showISSN{2220-9964}
\urldef\tempurl%
\url{https://doi.org/10.3390/ijgi7110427}
\showDOI{\tempurl}


\bibitem[Teitler et~al\mbox{.}(2008)]%
        {Teit08}
\bibfield{author}{\bibinfo{person}{Benjamin~E. Teitler},
  \bibinfo{person}{Michael~D. Lieberman}, \bibinfo{person}{Daniele Panozzo},
  \bibinfo{person}{Jagan Sankaranarayanan}, \bibinfo{person}{Hanan Samet},
  {and} \bibinfo{person}{Jon Sperling}.} \bibinfo{year}{2008}\natexlab{}.
\newblock \showarticletitle{NewsStand: A New View on News}. In
  \bibinfo{booktitle}{\emph{Proceedings of the 16th ACM SIGSPATIAL
  International Conference on Advances in Geographic Information Systems}}
  (Irvine, California) \emph{(\bibinfo{series}{GIS '08})}.
  \bibinfo{publisher}{Association for Computing Machinery},
  \bibinfo{address}{New York, NY, USA}, Article \bibinfo{articleno}{18},
  \bibinfo{numpages}{10}~pages.
\newblock
\showISBNx{9781605583235}
\urldef\tempurl%
\url{https://doi.org/10.1145/1463434.1463458}
\showDOI{\tempurl}


\bibitem[Times(2022)]%
        {toolNYT}
\bibfield{author}{\bibinfo{person}{The New~York Times}.}
  \bibinfo{year}{2022}\natexlab{}.
\newblock \bibinfo{title}{New York Times COVID-19 dashboard}.
\newblock
\newblock
\urldef\tempurl%
\url{https://www.nytimes.com/interactive/2021/world/covid-cases.html}
\showURL{%
Retrieved May 28, 2022 from \tempurl}


\bibitem[Turkay et~al\mbox{.}(2014)]%
        {Turk14}
\bibfield{author}{\bibinfo{person}{Cagatay Turkay}, \bibinfo{person}{Aidan
  Slingsby}, \bibinfo{person}{Helwig Hauser}, \bibinfo{person}{Jo Wood}, {and}
  \bibinfo{person}{Jason Dykes}.} \bibinfo{year}{2014}\natexlab{}.
\newblock \showarticletitle{Attribute Signatures: Dynamic Visual Summaries for
  Analyzing Multivariate Geographical Data}.
\newblock \bibinfo{journal}{\emph{IEEE Transactions on Visualization and
  Computer Graphics}} \bibinfo{volume}{20}, \bibinfo{number}{12}
  (\bibinfo{year}{2014}), \bibinfo{pages}{2033--2042}.
\newblock
\urldef\tempurl%
\url{https://doi.org/10.1109/TVCG.2014.2346265}
\showDOI{\tempurl}


\bibitem[University(2022)]%
        {toolJHU}
\bibfield{author}{\bibinfo{person}{Johns~Hopkins University}.}
  \bibinfo{year}{2022}\natexlab{}.
\newblock \bibinfo{title}{Johns Hopkins COVID-19 dashboard}.
\newblock
\newblock
\urldef\tempurl%
\url{https://coronavirus.jhu.edu/}
\showURL{%
Retrieved May 28, 2022 from \tempurl}


\bibitem[Łucjan(2016)]%
        {Lucj16}
\bibfield{author}{\bibinfo{person}{Kamila Łucjan}.}
  \bibinfo{year}{2016}\natexlab{}.
\newblock \showarticletitle{Perception of the contents of animated maps}.
\newblock \bibinfo{journal}{\emph{Polish Cartographical Review}}
  \bibinfo{volume}{48} (\bibinfo{date}{12} \bibinfo{year}{2016}).
\newblock
\urldef\tempurl%
\url{https://doi.org/10.1515/pcr-2016-0015}
\showDOI{\tempurl}


\end{thebibliography}
